\documentclass[]{iopart}
\usepackage{graphicx}

\def\gtorder{\mathrel{\raise.3ex\hbox{$>$}\mkern-14mu
 \lower0.6ex\hbox{$\sim$}}}
\def\ltorder{\mathrel{\raise.3ex\hbox{$<$}\mkern-14mu
 \lower0.6ex\hbox{$\sim$}}}

\newcommand{\be}{\begin{equation}}
\newcommand{\ee}{\end{equation}}
\newcommand{\bea}{\begin{eqnarray}}
\newcommand{\eea}{\end{eqnarray}}
\newcommand{\bra}[1]{\langle {#1} |}
\newcommand{\ket}[1]{| {#1} \rangle}

\def\>{{\rangle}}

\def\M{{\cal M}}

\def\F{{\cal F}}
\def\bp{{\bf p}}
\def\bP{{\bf P}}

\def\vk{{\vec k}}
\def\Tr{{\rm Tr}}

\def\half{{\textstyle{1\over 2}}}

\begin{document}

\title{ Neutron Structure Functions}

\author{J. Arrington, F. Coester, R. J. Holt, T.-S. H. Lee}
\address{Physics Division, Argonne National Laboratory, Argonne, Illinois
60439, USA}

\date{\today}

\begin{abstract}

Neutron structure functions can be extracted from proton and deuteron data
and a representation of the deuteron structure.  This procedure does not
require DIS approximations or quark structure assumptions.
We find that the results depend critically on properly accounting for the
$Q^2$ dependence of proton and deuteron data.  We interpolate the data to
fixed $Q^2$, and extract the ratio of
neutron to proton structure functions.  The extracted ratio decreases with
increasing $x$, up to $x \approx 0.9$, while there are no data available to 
constrain the behavior at larger $x$.

\end{abstract}

\maketitle

\section{Introduction}\label{sec:intro}
Traditionally, much of our information regarding the quark structure of the
nucleon has been gleaned from charged lepton scattering. If the one-photon
exchange approximation is sufficient, then inclusive structure functions of a
target nucleon or nucleus are well defined invariant functions of the cross
sections. The structure functions are dimensionless invariant functions of
the target four-momentum $P$ and the four-momentum transfer $q$. Thus nucleon
structure functions are functions of two invariants, {\it e.g.} the momentum
transfer, $Q^2=-q^2$, and the energy transfer, $\nu=P\cdot q /m$, where $m =
\sqrt{P^2}$ is the nucleon mass. The structure functions can also be taken as
functions of other invariants, {\it e.g.} $Q^2$ and 
$x$, where $x= Q^2/2P\cdot q$.  It follows from the definition of $x$ that
\be
x=1-{(P+q)^2-P^2\over 2P\cdot q} \leq 1\; ,
\ee
independent of any structure assumptions. For ``deep-inelastic'' scattering,
where
\be
{\nu^2\over Q^2}\equiv {(P\cdot q)^2\over m^2 Q^2}\equiv{Q^2\over 4 m^2 x^2}>>1 \;,
\label{eq:dis}
\ee
the parton model implies that the structure functions depend only on $x$
(``scaling''), with a weak logarithmic dependence on $Q^2$ (``scaling
violations'') due to QCD evolution of the parton distributions.

Since the discovery of scaling in deep inelastic electron--proton
scattering~\cite{bloom70, breidenbach69, bjorken67, bjorken69b,
feynman69}, there have
been a number of measurements~\cite{amaudruz92, arneodo97, benvenuti89,
aubert87, whitlow90} of the proton structure function. The relation of the
proton structure function to the quark structure of the nucleon is model
dependent. At medium energies and large $x$, e.g. $Q^2=10$~GeV$^2$ and $x=0.9$,
yielding $Q^2/4m^2 x^2 = 3.5$, the limit of (\ref{eq:dis}) is not satisfied
and the scattering is not ``deep inelastic''.

To test models of the quark structure of nucleons it is important to have
experimental results for both protons and neutrons. The nucleon structure
functions for $x>0.4$ are particularly important for constraining models.
Since neutron targets are not practical, experimental neutron structure
functions must be extracted from deuteron and proton scattering data, without
any quark assumptions. This requires consistent Poincar\'e covariant
representations of both the deuteron states and the current-density
operators.

Realistic nucleon-nucleon potentials are derived from Lagrangians which
determine two-nucleon Green functions, and the interaction dependent exchange
currents~\cite{gross87,coester94}. Different truncations, regularizations and
Lorentz invariant constraints are used for the construction of effective
potentials. Parameters are adjusted to accurately fit both deuteron properties
and nucleon-nucleon scattering data. These potentials define two-nucleon
rest-energy operators (mass operators), which are both Lorentz invariant and
Galilean invariant. The difference between the binding energy $M_d-2m$ and
$(M_d^2-4m^2)/4m$ is a negligible relativistic effect. Spin operators are
invariant under Galilean boosts, and undergo Wigner rotations under Lorentz
boosts. This difference is the source of important relativistic effects. It
does not, however, affect inclusive structure functions of unpolarized
targets.  If these mass operators fit the same scattering and bound state data
then they are related by unitary transformations. Such unitary transformations
modify the separation of current density operators into one- and two-body
operators.

Given the dynamics specified by an invariant mass operator, unitary
representations of the full Poincar\'e group obtain readily by specification
of the generators as functions of the mass operator. The generators of a
kinematic subgroup are independent of the mass operator. This choice of
kinematics which affects the representation of the current density operators,
can be exploited to simplify the relations of deuteron structure functions
to nucleon structure functions. The identification of single-nucleon currents
is representation dependent. With null-plane kinematics~\cite{coester92} the
representations of Lorentz transformations that leave a null vector $n=\{1,
-\hat n\}$ invariant are independent of the dynamics and only the momentum
component $P^-=\ell\cdot P,\;\;\ell=\{1,\hat n\}$ depends on the mass
operator. The null vector $n$ can be chosen such that $Q^+=n\cdot q=0$. Thus
$Q^2=Q_\perp^2-Q^+Q^-=Q_\perp^2$ is independent of the dynamics.
Kinematic Lorentz transformations which change $P_\perp$ do not affect
$Q^+=0$.  With this form of kinematics the impulse assumption for the relevant
components of the current density is consistent with current conservation and
Lorentz covariance.

This form of kinematics permits a Lorentz invariant convolution relation of
the nucleon and deuteron structure functions, which does not require the
deep-inelastic approximation, and is independent of the mass spectrum of the
final state.  Such ``smearing'' is $Q^2$ dependent but converges to the
familiar $Q^2$ independent convolution when $Q^2/\nu^2$ is negligible. It is
convenient to choose $P^+=2m$ and $P_\perp$ such that
\be
P\cdot q = -\vec P_\perp \cdot \vec Q_\perp\; ,
\quad \nu=-{\vec P_\perp \cdot \vec Q_\perp\over m} \; .
\label{FRAME}
\ee
The values of invariants are, of course, independent of such a choice.

In this paper we extract the neutron structure function from proton and
deuteron data taking the impulse assumption and an effective two-nucleon mass
operator.  The main focus is on the consequences of systematic uncertainties
and the $Q^2$ dependence in the proton and deuteron data. A systematic
exploration of the uncertainties associated with using different models of
deuteron structure is beyond the scope of this paper.

\section{Cross Sections and Inclusive Structure Functions}

For inclusive electron scattering from an unpolarized target nucleus the
Lorentz invariant inclusive cross section can, in the one-photon exchange
approximation, be expressed in the form
\be
{d\sigma\over d\Omega d E'}=
{ 4 \alpha^2 {E'}^2\cos^2\theta/2 \over Q^4 M} F^{\mu\nu}(P,q)f_{\mu\nu}\; ,
\ee
where $M=\sqrt{P^2}$ is the target mass and the electron current tensor,
\be
f_{\mu\nu}= \half\left({k_\mu k'_\nu+k'_\mu k_\nu\over (k\cdot k')} -g_{\mu\nu}\right)\;,
\ee
is a bilinear function of the initial and final electron momenta $k$ and $k'$.
It follows from Lorentz covariance and current conservation that the current
tensor $F^{\mu\nu}(P,q)$ is a function of two invariant functions,
$F_1(\nu,Q^2)$ and $F_2(\nu,Q^2)$,
\be
F^{\mu\nu}(P,q)= \Bigl({q^\mu q^\nu\over q^2}-g^{\mu\nu}\Bigr) F_1(\nu,Q^2)
+{\tilde P^\mu\, \tilde P^\nu\over P\cdot q}F_2(\nu,Q^2)\;,
\ee
where
\be
\tilde P= P-{P\cdot q\over q^2}\,q\;.
\ee
In the frame specified by (\ref{FRAME}) and $Q_\perp=\{\sqrt{Q^2},0\}$, the
invariant functions are proportional to single components $F^{22}$ and
$F^{++}= n_\mu n_\nu F^{\mu\nu}$ of the current tensor
\bea
F_1(\nu,Q^2)=F^{22}(P,q)\; , \quad
F_2(\nu,Q^2)= (P\cdot q) {F^{++}(P,q)\over {P^+}^2}\; .
\label{FF}
\eea
Since $q^\mu f_{\mu\nu} =0$ it follows from these definitions that
\bea
{d\sigma\over d\Omega d E'}=
{ 4\alpha^2 {E'}^2 M\cos^2\theta/2 \over Q^4 (P\cdot q) }
\left[F_2(\nu, Q^2)+{2 P\cdot q\; \tan^2\theta/2 \over M^2}F_1(\nu,Q^2)
\right],\cr
\eea
and with the ratio of longitudinal and transverse cross sections
$R=\sigma_L/\sigma_T$,
\be
1+ R(\nu,Q^2)={M^2\over P\cdot q}{F_2(\nu,Q^2)\over F_1(\nu,Q^2)}
\left(1+{\nu^2\over Q^2}\right) \;,
\ee
\bea
{d\sigma\over d\Omega d E'}=
{ 4\alpha^2 {E'}^2 M\cos^2\theta/2 \over Q^4 P\cdot q }
F_2\left[1 +{2 (Q^2+\nu^2)\over (1+R)Q^2}\; \tan^2\theta/2 \right].
\eea
The relation of structure functions to cross sections does not involve any
assumptions about the target structure.

\section{Proton and Deuteron Data}

To extract neutron structure functions using convolution relations we need
proton and deuteron structure functions at a fixed value of $Q^2$. To minimize
systematic uncertainties, we use a fit to the proton structure function,
$F_{2p}$, from a recent global analysis and direct measurements of the
\textit{ratios} of deuteron to proton structure function, $R_{dp} =
F_{2d}/F_{2p}$.  Therefore, only the ratios need to be interpolated to fixed
values of $Q^2$, minimizing the size of these correction.  In addition, we
minimize sensitivity to normalization uncertainties between data sets by using
only measurement of $F_{2d}$ and $F_{2p}$ taken from the same experiment when
forming the deuteron-to-proton ratios.

For $F_{2p}(x,Q^2)$, we use the fit from~\cite{christy_priv}, evaluated at
$Q_0^2=12$~GeV$^2$, which provides parameterizations for $F_{1p}$ and
$F_{2p}$. The parameters are fit to a large body of data~\cite{arneodo97,
benvenuti89, aubert85, whitlow92, tao95, tvaskis07}. The $Q^2$ dependence comes
from the fit to the data, rather than from a specific model of scaling
violations.  We compared the fit to the $F_{2p}$ extraction from several SLAC
measurements~\cite{whitlow92, whitlowphd} and BCDMS~\cite{benvenuti89}. For
both experiments, the data are taken from the analysis of
Whitlow~\cite{whitlow92, whitlowphd}, where a relative normalization factor
between different experiments is determined and all data are extracted using a
common parameterization for $R=\sigma_L/\sigma_T$. The fit is consistent with
the data within the quoted uncertainties, yielding $\chi^2_\nu=0.86$ when
compared to all data for $6 < Q^2 < 40$~GeV$^2$.

For the ratio $R_{dp}(x,Q^2)$, we take the measurements from
SLAC~\cite{whitlow92}, BCDMS~\cite{benvenuti89, benvenuti90a}, and
NMC~\cite{arneodo97}.  The SLAC data are from a reanalysis of several
experiments, and included a fit to the relative normalizations for the
different SLAC experiments.  While the SLAC, BCDMS, and NMC used different
models for $R=\sigma_L/\sigma_T$, this has no impact on the structure function
ratios, as all of the extractions used identical values for $R$ is the proton
and deuteron, as suggested by world's data~\cite{whitlow90, tvaskis07}.

In combining all of the $R_{dp}(x,Q^2)$ measurements, we have to account for
both statistical uncertainties and the relative normalization of the data
sets.  The uncertainties are broken up into statistical, correlated
systematic, and normalization uncertainties.  We take the statistical
uncertainties to be the combination of the counting statistics and the
\textit{uncorrelated} systematic uncertainties.  The correlated systematic
uncertainties come mainly from the relative normalization of the different
data sets, which yields uncertainties that are typically highly correlated
between neighboring $x$ points. The detailed analysis~\cite{whitlowphd}
yielded a 1\% overall normalization unceratinty, with smaller relative
normalization uncertainties. In combining the results with the BCDMS and SLAC
measurements, we also fit a relative normalization offset between the
experiments.  In the final analysis, we scale the SLAC ratios down by 0.8\%,
and the BCDMS and NMC results up by 1.9\% and 0.1\% respectively, in all cases
within the systematic uncertainties quoted for the ratios.  We take a 1\%
overall normalization uncertainty on the ratio, and use uncertainties in the
relative normalizations of 0.5\% for NMC, 1.0\% for BCDMS, and 0.4--1.1\% for
the SLAC experiments, as determined in Ref.~\cite{whitlowphd}.

Because the proton structure function parameterization does not include a
contribution associated with elastic scattering, the contribution from
quasielastic scattering in the deuteron would not be reproduced. The
kinematics are chosen such that this contribution is extremely small. A fit
to quasielastic scattering data, made to estimate quasielastic and inelastic
contributions in the analysis of large $x$ inclusive scattering
data~\cite{e02019}, indicates that the quasielastic contribution to the
$R_{dp}$ ratio is negligible, with a contribution that is at most 0.3\% (for
the largest $x$ values), and is always much less than the experimental
uncertainties.

There are several points in the results of~\cite{whitlowphd} where the quoted
values for $F_{2d}$ and $F_{2p}$ do not match the value given for the ratio
$F_{2d}/F_{2p}$. This usually corresponds to cases where the $x$ or $Q^2$
values of the hydrogen and deuterium data did not match exactly, and
corrections had to be applied to form ratios at identical kinematics. This
correction can be large, especially at large $x$ where $F_{2p}$ drops rapidly
with increasing $x$.  This correction is model-dependent, and can introduce
a significant uncertainty if the correction is large.  In addition, there are
a few cases where the discrepancy is too large to be explained by the
interpolation (up to 25\% in one case), suggesting an error in either the
quoted structure functions or the ratio.  To avoid introducing possible bias
due to such errors and to minimize the model-dependent corrections associated
with large interpolations, we eliminated points where the quoted value for the
ratio is more than 2\% from the value calculated from the quoted $F_{2d}$ and
$F_{2p}$ values.  The impact of this cut is negligible except at very large
$x$, where the effect is still small compared to the experimental
uncertainties.  As the final result is insensitive to the exact value of the
cut, we do not include any additional uncertainty associated with this cut.

This procedure gives us a large number of individual values of the ratio
$R_{dp}(x,Q^2)$, which need to be interpolated to a fixed $Q^2$ value,
$Q_0^2$, and binned in $x$. We choose $Q_0^2=12$~GeV$^2$, as this is the
average $Q^2$ value of the large-$x$ data, and limit the overall $Q^2$ range
so that the average $Q^2$ in a given bin is within a factor of two of $Q_0^2$,
to minimize the size of the interpolation correction.

It has been observed that the $Q^2$ dependences of the proton and nuclear
structure functions~\cite{filippone92, niculescu00b, arrington01,
melnitchouk05} and structure function ratios~\cite{arrington06a} are much
smaller when taking fixed values of $\xi = 2x/(1+\sqrt{1+4m^2x^2/Q^2})$
rather than fixed $x$. Thus, we fit the $Q^2$ dependence at fixed $\xi$
over the full $Q^2$ range of the data, from 3--230~GeV$^2$, to extract the
$Q^2$ dependence. We find $dR_{dp}/d(\ln(Q^2)) \approx -0.013$, with little
$x$ dependence, except for $x \ltorder 0.1$ and $x \gtorder 0.7$, where the
$Q^2$ dependence appears to be smaller, but is not well measured.  This
$Q^2$ dependence is used to interpolate the individual data points to $Q_0^2$.

For the final analysis, we use only data with $W^2>3$~GeV$^2$ and
$6<Q^2<40$~GeV$^2$, yielding a maximum interpolation correction of $<$1\%, and
an RMS correction of 0.4\%. In the regions where the $Q^2$-dependence is not
well measured, this correction is small compared to the uncertainties, and the
associated uncertainty is negligible. After all of the cuts, we have 450
$R_{dp}$ measurements, all corrected to $Q_0^2$. We then bin the data in $x$ to
extract a set of points for $R_{dp}(x,Q_0^2)$. We combine the data points by
taking the statistics-weighted average of the individual points, and the final
systematic uncertainty is taken to be the statistics-weighted average of the
individual systematic uncertainties.

\section{Nuclear Target Structure}

The current tensor $F^{\mu\nu}$ of the target,
\bea
F^{\mu\nu}(P,q)= \Tr \left( \F^{\mu\nu} (q)\rho_P \right)\; , \quad
\F^{\mu\nu} (q)=I^\mu (0)\tilde I^\nu(q)\; ,
\eea
is a functional of the current-density operator
\bea
I^\mu(x)=e^{\imath P\cdot x}I^\mu(0)e^{-\imath P\cdot x}\; , \quad
\tilde I^\nu(q)={1\over (2\pi)^4}\int d^4q e^{-\imath q\cdot x} I^\mu(x)\; ,
\eea
and the target density of a nucleus with spin $j$
\be
\rho_P= {1\over 2 j +1}\sum_{\sigma =-j}^{+j}\ket{\sigma,P}\bra{ P ,\sigma}\;.
\ee

Lorentz transformations, $\Lambda$, are represented by unitary operators
$U(\Lambda)$. The current density operator and the target density transform
according to
\bea
U^\dagger(\Lambda)\F^{\mu\nu}(q) U(\Lambda)=
\Lambda^\mu\,_{\mu '} \Lambda^\nu\,_{\nu '}
\F^{\mu'\nu'} (\Lambda^{-1}q)\; , \quad\cr
U^\dagger(\Lambda)\rho_P U(\Lambda) =\rho_{_{\Lambda P}}\; .
\eea
The wave function representing the target state which generates that
density may or may not be ``manifestly'' covariant. If the covariance is not
``manifest'' there always exist transformations to ``manifestly'' covariant
representations, but there is no substantial significance to ``manifest''
covariance. Representations of target structure that emphasize the importance
of ``manifest'' covariance intend to emphasize features of quantum field theory
which affect the representation of the target structure.

Assuming the only constituents of the deuteron are a proton and a neutron, the
deuteron states $\ket{\sigma, P}$ may be represented by functions
$\Psi_{\bP,M_d,\sigma} (\bp_p,\lambda_p,\bp_n,\lambda_n)$, of the null-plane
momenta and the null-plane spins. The null-plane spins are invariant under the
kinematic Lorentz transformations that leave the null-plane invariant. These
functions are eigenfunctions of an invariant mass operator, $\M$ and the
four-momentum operator $P$:
\be
P_\perp=p_{p\perp}+p_{n\perp}\; , \quad
P^+=p_p^++p_n^+\; , \quad
P^-={\M^2+P_\perp^2\over P^+} \;.
\ee
The associated representation of the current tensor operator, \\
$\bra{\bp_n',\lambda_n',\bp_p',\lambda_p'} \F^{\alpha\alpha}(q)
\ket{\bp_p,\lambda_p,\bp_n, \lambda_n}$, is the sum of the impulse term,
\bea
&&\bra{\bp_n',\lambda_n',\bp_p',\lambda_p'}
\F_{imp}^{\alpha\alpha}(q)\ket{\bp_p,\lambda_p,\bp_n,\lambda_n} \cr
&&=\delta(\bp_n'-\bp_n)\delta_{\lambda_n',\lambda_n}
\bra{\bp_p',\lambda_p'}\F_p^{\alpha\alpha}(q)\ket{\bp_p,\lambda_p} \cr
&&+ \delta(\bp_p'-\bp_p)\delta_{\lambda_p',\lambda_p}
\bra{\bp_n',\lambda_n'}\F_n^{\alpha\alpha}(q)\ket{\bp_n,\lambda_n} \; ,
\eea
and the remainder, $\F_{int}^{\alpha\alpha}(q)$, which vanishes for large
separation of the nucleons. The impulse term is essential for the relation
of nucleon and deuteron structure functions. The representation of
$\F_{int}^{\alpha\alpha}(q)$ is clearly model dependent. Any unitary
transformation of $\rho_p$ which does not affect the N-N data~\cite{bogner07}
modifies $\F_{int}^{\alpha\alpha}(q)$ and thus any conclusion relating
nucleon and deuteron structure functions. For $\alpha= +$ and $2$, the impulse
assumption $\F_{int}^{\alpha\alpha}=0$ is consistent with the requirements of
Poincar\'e covariance and current conservation.

It follows from the relation (\ref{FF}) and the impulse assumption that the
deuteron structure function is related to the proton and neutron structure
functions,
\bea
F_{2d}(\nu, Q^2)= \half (\overline F_{2p}(\nu,Q^2)+ \overline F_{2n}(\nu,Q^2))
\label{eq:f2d}\; , \cr\cr
\overline F_{2N}(\nu,Q^2)= \int {dp^+\over p^+} \int d^2p_{\perp}
{{p^+}^2 \over {P^+}^2} \rho_P(\bp){\nu \over \nu_N}
F_{2N}(\nu_N, Q^2)\; ,
\label{conv}
\eea
where $\nu_N= -(\vec p_{\perp}\cdot \vec Q) / m$.
No deep inelastic approximations are involved and there are no target mass
effects. Target mass effects are artifacts of inconvenient representations of
the deuteron states and the current density operators~\cite{oelfke90}.

Nucleon momentum densities $\rho_P(\bp)$ obtained from a deuteron wave
function satisfy the normalization condition
\be
\int {dp^+\over p^+}\int d^2 p_\perp \rho_P(\bp)=1 \; .
\ee
With the definitions $z= 2 p^+/P^+$ and $\vec k_\perp=\vec p_\perp-\half z
\vec P_\perp$ the density $\rho(z,\vk_\perp)=\rho_P(\bp)$ satisfies
\be
\int _0^2{dz\over z} \int d^2 k_\perp \rho(z,k_\perp)=1\; .
\ee

Conventional nuclear ground-state wave functions are eigenfunctions of a mass
operator which are easily related to eigenfunctions of the four-momentum
operator. In particular a deuteron wave function which satisfies
\be
\left(V + \vk^2/ m \right)\varphi_d= \varphi_d\, E_d \; ,
\ee
where $V$ is any ``realistic'' nucleon-nucleon potential, and is an
eigenfunction of an invariant two-nucleon mass operator,
\be
(4 \omega^2+4m V)\varphi_d(\vk)=\varphi_d(\vk)\, M_d^2
\left(1- {E_d^2 / M_d^2}\right)\; , \quad
\ee
with $\omega=\sqrt{\vk\,^2+m^2}$ and
\bea
\varphi_{d \sigma}(\vk,\sigma_1,\sigma_2)=u(|\vk|)
(\half,\half,\sigma_1,\sigma_2|1,\sigma)\cr\cr
+\sum_{m,\sigma'}w(|\vk|)Y_2^n(\hat k)(\half,\half,\sigma_1,\sigma_2|1,\sigma')
(2,1,n,\sigma'|1,\sigma)\; .
\eea
The ratio $E_d^2/ M_d^2\approx 10^{-6}$ is negligible. The invariant
momentum density is
\be
\rho(\vk)= {1\over4\pi}\left(|u(|\vk|)|^2+ |w(|\vk|)|^2\right)\;,\quad
\int d ^3 k\, \rho(\vk) =1\; .
\ee
The density $ \rho(z,k_\perp)\equiv\rho(z,-k_\perp)\equiv \rho(2-z,k_\perp) $ is
related to $\rho(\vk)\equiv \rho(-\vk)$ by
\be
\rho(z,k_\perp)= z \int_{-\infty}^{+\infty}dk_z \rho(\vk)
\delta (z-1-k_z / \omega) \; .
\ee

With these variable transformations the convolution relation of the structure
functions, (\ref{conv}) takes the form
\bea
\overline F_{2N}(\nu,Q^2)
&=&\int_0^2 dz \int d ^2k_\perp {1\over z-\Delta_N} \rho(z,k_\perp ) \cr\cr
&&\times F_{2N}\left(\nu_N,Q^2 \right) \theta\left(m\nu_N-\half Q^2\right)\; ,
\label{CON}
\eea
with
\be
\Delta_N = {k_\perp\cdot Q_\perp\over m\nu} = {2\vec k\cdot \vec Q_\perp\over Q^2 } x
  \; , \quad
\nu_N = \nu(z -\Delta_N)\; .
\ee

The Melosh rotations which relate null-plane spins to canonical spins do not
affect these relations. Whether in practice the structure functions are
parameterized as functions of $\nu$ and $Q^2$, $x$ and $Q^2$, or $\xi$ and
$Q^2$, is a matter of convenience. For interpolations and averaging of nucleon
data, $F_2(\xi,Q^2)$ has been used effectively. In the convolution, there
is no particular justification for the use of of $\xi$, and $x$ is a more
convenient choice. The kinematic constraint on the $\nu$ and $Q^2$ dependence
of the structure functions implies $0 \leq x \leq 1$ for the nucleons and
$0 \leq x \leq 2$ for the deuteron, making it convenient to represent structure
functions as functions $F_2(x,Q^2)$. In that representation the convolution
relation, (\ref{CON}), takes the following form:
\bea
\overline F_{2N}(x,Q^2)
= \int d ^3k{z(\vk)\over z(\vk)-\Delta_N} \rho(|\vk| )
F_{2N}\left(x_N,Q^2 \right) \theta\left(1-x_N \right) \; ,
\label{CONX}
\eea
where
\be
z(\vk)=1+ k_z / \omega\; ,\quad
x_N = \frac{x}{z-\Delta_N}\; .
\ee
Note that in the case of the deuteron, $\vec k_p = - \vec k_n$, so $z_{p,n}=1
+ k_z^{p,n} / \omega = 1 \pm k_z^p/\omega$ and $\Delta_p = - \Delta_n$.

The main contributions are from values of $z$ in the neighborhood of one
(between 0.8 and 1.2), and $\Delta_N$ is small compared to one, but not
negligible. In the limit of $\Delta_N \to 0$, the relation (\ref{CONX}) reduces
to the familiar deep-inelastic convolution:
\bea
\overline F_{2N}(x,Q^2)&=& \int dz f(z) F_{2N}\left( {x\over z},Q^2\right)\theta(z-x)\; , 
\cr\cr
f(z) &=&\ \int d^2k_\perp \rho(z ,k_\perp)\; .
\eea

\section{Extraction of $ F_{2n}$ from Proton and Deuteron Data}\label{sec:extraction}

\subsection{Extraction procedure}

Given the fit for $F_{2p}$ from~\cite{christy_priv}, and the deuteron
momentum density calculated from the CD-Bonn potential~\cite{machleidt01}, we
use (\ref{CONX}) to calculate $\overline F_{2p}$. With a parameterization
of $F_{2n}$ we obtain $\overline F_{2n}$ and thus values of $F_{2d}$, which are
compared to the data. Because the ratio $R_{dp}=F_{2d}/F_{2p}$ is measured
with smaller uncertainties than the absolute structure functions, we compare
the calculated values of $R_{dp}$ with the measured values to fit the
parameterization of the neutron structure function. We parameterize $R_{np}$
at $Q^2=12$~GeV$^2$ and then use MINUIT~\cite{james75} to find the fit to
$R_{np}$ that best reproduces the measured deuteron to proton ratio.

At fixed $Q^2$ it is a matter of convenience whether one parameterizes
$R_{np}=F_{2n}/F_{2p}$ as a function of $x$ or $\xi$. We chose to
parameterize $R_{np}$ as a function of $\xi$, with the expectation that
$F_{2n}(\xi, Q^2)$ will have a smaller $Q^2$ dependence than $F_{2n}(x,Q^2)$,
as is the case for the proton and deuteron.
We parameterize the $R_{np}(\xi,Q_0^2)$ using the following form,
\bea
R_{np}(\xi) = (p_1+p_2 \xi) + p_3 \exp{(-p_4 \xi)} \nonumber \\
+ p_5 \exp{(-p_6 (1-\xi))}  + p_7 [\max{(0,\xi-p_8)}]^2\; ,
\label{eq:npfit}
\eea
where the linear and first exponential terms are the dominant pieces at low
$x$, and the quadratic and exponential terms provide flexibility at large $x$.
At $x=0$, $R_{np} \approx p_1 + p_3$, as the other terms are negligible.
The initial fits gave values of $R_{np}$ consistent with unity at
$x=0$, and so we apply the constraint that the ratio should go to one, and
take $p_3=1-p_1$ for all subsequent analysis. The last two terms yield
deviation from the linear behavior at large $x$; by including two terms,
we allow the possibility of significant cancellation between these terms,
yielding flexibility for a modification of the shape at intermediate $x$
values while still reproducing the data at large $x$.

\begin{figure}[htb]
\begin{center}
\includegraphics*[width=3.7in]{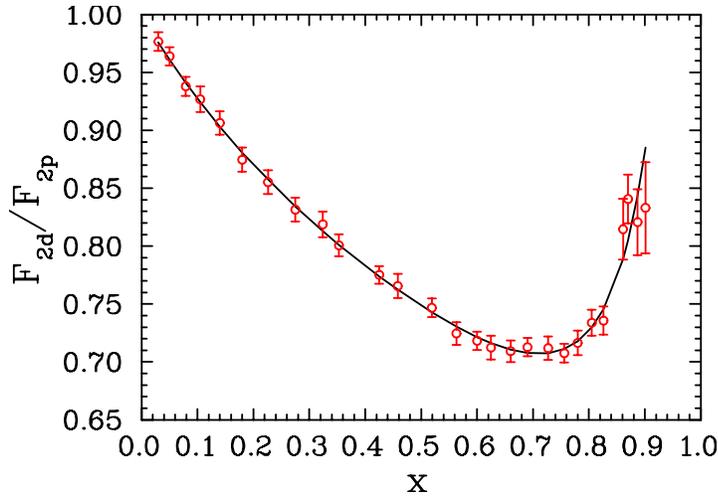}
\end{center}
\caption{Extracted $F_{2d}/F_{2p}$ for $Q^2$=12~GeV$^2$ (points), along with
the calculation taking $R_{np}$ from the best fit using the parameters in
table~\ref{tab:fitpar}.}
\label{fig:dprat}
\end{figure}

\begin{table}[ht]
\caption{Fit parameters for the best fit to $R_{np}$ using the parameterization
from (\ref{eq:npfit}). Note that $p_3$ is not varied in the fit, and is taken
to be $(1-p_1)$.
\label{tab:fitpar}}
\begin{indented}
\item[]\begin{tabular}{|c|r|c|r|}
\hline
Parameter & value~~& Parameter & value~~ \\
\hline
$p_1$  &   0.816~~ & $p_5$ & --0.034~~\\
$p_2$  & --0.661~~ & $p_6$ &   8.714~~\\
($p_3$)&   0.184~~ & $p_7$ & --0.072~~\\
$p_4$  &   5.509~~ & $p_8$ &   0.450~~\\ \hline
\end{tabular}
\end{indented}
\end{table}

\begin{figure}[htb]
\begin{center}
\includegraphics*[width=3.7in]{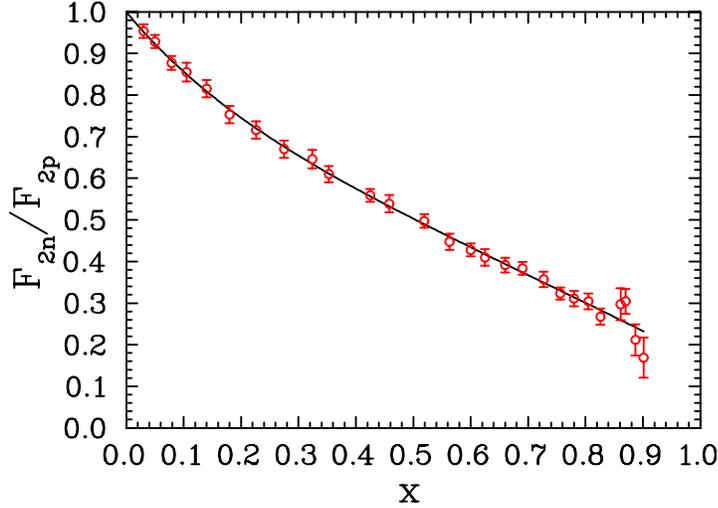}
\end{center}
\caption{Extracted $F_{2n}/F_{2p}$ for $Q^2$=12~GeV$^2$. The data points
represent values extracted from $R_{dp}$ measurements according
to~(\ref{eq:smearing2}) while the curve is given by (\ref{eq:npfit}) with the
parameters in table~\ref{tab:fitpar}.}
\label{fig:nprat}
\end{figure}

Figure~\ref{fig:dprat} shows the $R_{dp}$ data interpolated to
$Q_0^2$=12~GeV$^2$. The uncertainties shown are the combined statistical and
uncorrelated systematic uncertainties. There is also an estimated 1\%
normalization uncertainty~\cite{whitlow92}. The curve in
figure~\ref{fig:dprat} shows the calculated $R_{dp}(x,Q_0^2)$ for the best fit
to $R_{np}$ (the parameters of the fit are given in table~\ref{tab:fitpar}). 
Because the uncertainties are almost evenly split between statistical and
correlated systematic except at the largest $x$ values, we expect a reduced
chi-squared of less then unity. Including the contribution from the
normalization paremeters, the fit gives $\chi^2=14.6$ for 17 degrees of
freedom (27 data points, 7 parameters in the $x$-dependence, and 3
normalization factors).

Figure~\ref{fig:nprat} shows the fit for $R_{np}$, along with the values of
$R_{np}$ extracted from the $R_{dp}$ measurements. In terms of the smearing
ratios, $S_N(x,Q^2) = \overline F_{2N}(x,Q^2) / F_{2N}(x,Q^2)$, we can
write (\ref{eq:f2d}) as:
\be
F_{2d} = \half ( S_p F_{2p} + S_n F_{2n} )\; ,
\label{eq:smearing}
\ee
which allows us to extract $R_{np}$ using the simple expression:
\be
R_{np} = ( 2 R_{dp} - S_p ) / S_n .
\label{eq:smearing2}
\ee
The advantage of extracting individual $R_{np}$ points using these smearing
ratios, calculated from the extracted $F_{2n}(x,Q_0^2)$, is that it provides an
estimate of the uncertainty in $R_{np}$ coming from the uncertainty in the
$R_{dp}$ measurements. While this is convenient, and consistent with the way
previous extractions have been presented, it does not take into account that a
given value of $R_{dp}(x,Q^2)$ depends on the proton and neutron structure
functions over a range in $x$.  This will be discussed in
section~\ref{sec:systematics}.

\subsection{Systematic uncertainties}\label{sec:systematics}

The extracted $R_{np}(x)$ values shown in figure~\ref{fig:nprat} include the
uncertainties associated with the $R_{dp}$ data, but not the uncertainties
associated with the deuteron model and the impulse assumption or systematic
uncertainties associated with the extraction procedure. These uncertainties
tend to be important at large $x$, and will tend to have a highly correlated
impact on the high-$x$ results, and thus the extrapolation to $x=1$.
Because the systematic uncertainties were observed to grow with $x$, and yield
highly correlated corrections among the high-$x$ points, we make estimates
of the uncertainties and then fit these to an exponential, to
provide a simple parameterization of the correlated systematic error band.

The largest systematic at all $x$ values was due to the 1\% overall
normalization uncertainty on the global $R_{dp}$ measurements.  Shifting the
deuterium ratios by 1\% yields an overall absolute shift in $R_{np}$ of
approximately 0.014, plus an additional contribution at large $x$ values. The
large $x$ contribution is well described by $\delta R_{np} = 0.079
\exp{(-12(1-x))}$.

The result is also sensitive to the parameterization of the proton
structure function at large $x$. We vary the high-$x$ behavior of
$F_{2p}(x,Q^2)$, introducing a 10\% change in the falloff at large $x$,
which is large enough that the $F_{2p}$ fit starts to become
inconsistent with the SLAC measurements.  This yields a change of
$\delta R_{np} = 0.06 \exp{(-12(1-x))}$, where we have used the same
$x$ dependence as for the normalization systematic, as the behavior is
very similar.

\begin{figure}[htb]
\begin{center}
\includegraphics*[width=3.7in]{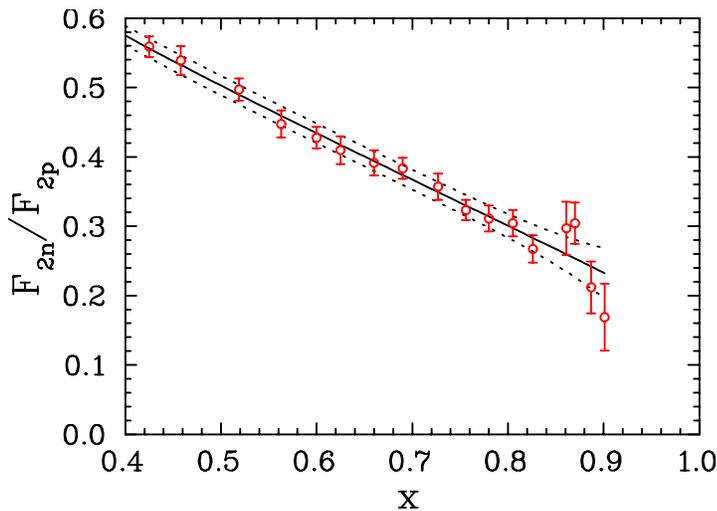}
\end{center}
\caption{Extracted value of $R_{np}$, along with the best fit (solid line) and
systematic uncertainty bands (dotted lines).}
\label{fig:syserr}
\end{figure}

\begin{table}[ht]
\caption{Extracted values and uncertainties for $R_{dp}(x)$ and $R_{np}(x)$ at
$Q^2$=12~GeV$^2$.  Errors labeled ``stat'' are the combined statistical and
uncorrelated systematic uncertainties.  Uncertainties labeled ``sys'' are 
the correlated systematic uncertainties coming from the relative normalization
between different data sets and are correlated.  The final column is the
uncertainty in the extracion of $R_{np}$ due to the systematic uncertainties
in the extraction procedure discussed in Sec.~\ref{sec:systematics}.  Not
included is the overall 1\% normalization uncertainty in $R_{dp}$ which 
corresponds to an overall shift of $\pm$0.014 in $R_{np}$.
\label{tab:ratio}}
\begin{indented}
\item[]\begin{tabular}{|c||c|c|c||c|c|c|c|}
\hline
$x$ & $R_{dp}$	& $\delta R_{dp}^{stat}$ & $\delta R_{dp}^{sys}$\dag &
$R_{np}$	& $\delta R_{np}^{stat}$ & $\delta R_{np}^{sys}$\ddag &
$\delta R_{np}^{ext}$ \\
\hline
 0.030 & 0.976 & 0.006 & 0.005 & 0.954 & 0.013 & 0.010 & 0.000 \\
 0.050 & 0.964 & 0.006 & 0.005 & 0.929 & 0.013 & 0.010 & 0.000 \\
 0.079 & 0.938 & 0.006 & 0.005 & 0.877 & 0.012 & 0.011 & 0.000 \\
 0.105 & 0.927 & 0.008 & 0.008 & 0.855 & 0.016 & 0.015 & 0.000 \\
 0.140 & 0.906 & 0.008 & 0.007 & 0.815 & 0.015 & 0.014 & 0.000 \\ \hline
 0.180 & 0.875 & 0.008 & 0.007 & 0.753 & 0.015 & 0.014 & 0.000 \\
 0.226 & 0.855 & 0.008 & 0.007 & 0.716 & 0.015 & 0.014 & 0.000 \\
 0.275 & 0.831 & 0.008 & 0.007 & 0.670 & 0.015 & 0.014 & 0.000 \\
 0.324 & 0.819 & 0.010 & 0.006 & 0.646 & 0.019 & 0.011 & 0.000 \\
 0.353 & 0.801 & 0.007 & 0.007 & 0.610 & 0.014 & 0.013 & 0.000 \\ \hline
 0.425 & 0.775 & 0.006 & 0.005 & 0.559 & 0.012 & 0.009 & 0.000 \\
 0.458 & 0.766 & 0.008 & 0.007 & 0.539 & 0.016 & 0.014 & 0.000 \\
 0.519 & 0.747 & 0.006 & 0.005 & 0.497 & 0.012 & 0.010 & 0.000 \\
 0.563 & 0.725 & 0.007 & 0.006 & 0.447 & 0.015 & 0.013 & 0.001 \\
 0.600 & 0.718 & 0.006 & 0.005 & 0.428 & 0.012 & 0.010 & 0.001 \\ \hline
 0.625 & 0.712 & 0.008 & 0.006 & 0.410 & 0.016 & 0.012 & 0.001 \\
 0.660 & 0.709 & 0.007 & 0.006 & 0.391 & 0.014 & 0.012 & 0.002 \\
 0.690 & 0.713 & 0.006 & 0.005 & 0.383 & 0.012 & 0.009 & 0.003 \\
 0.727 & 0.712 & 0.008 & 0.006 & 0.357 & 0.015 & 0.012 & 0.004 \\
 0.756 & 0.707 & 0.007 & 0.004 & 0.323 & 0.012 & 0.008 & 0.006 \\ \hline
 0.780 & 0.716 & 0.009 & 0.006 & 0.311 & 0.015 & 0.011 & 0.008 \\
 0.805 & 0.734 & 0.009 & 0.006 & 0.304 & 0.016 & 0.010 & 0.011 \\
 0.826 & 0.736 & 0.011 & 0.005 & 0.267 & 0.018 & 0.009 & 0.014 \\
 0.861 & 0.815 & 0.025 & 0.008 & 0.297 & 0.037 & 0.012 & 0.021 \\
 0.870 & 0.841 & 0.020 & 0.007 & 0.304 & 0.028 & 0.011 & 0.023 \\ \hline
 0.887 & 0.821 & 0.027 & 0.009 & 0.212 & 0.036 & 0.012 & 0.028 \\
 0.901 & 0.833 & 0.038 & 0.009 & 0.169 & 0.047 & 0.011 & 0.034 \\ \hline
\multicolumn{8}{|l|}{\dag Additional 1\% scale uncertainty is not shown.}\\ \hline
\multicolumn{8}{|l|}{\ddag Additional scale uncertainty of 0.014 is not shown.}\\
\hline
\end{tabular}
\end{indented}
\end{table}

Finally, the uncertainties shown on the $R_{np}$ data point come only
from the statistical and systematic uncertainties on the $R_{dp}$; there
is no contribution associated with the uncertainty in the factor
$S_n$, which depends on the extracted $F_{2n}$. To account for this,
we vary the parameters in the fit to $R_{np}$ and calculate $R_{dp}$. We
compare the extracted $R_{dp}$ to the data, and take the range of fits that
increase the total $\chi^2$ by one as the range of good $R_{np}$ fits.
For each of these, we recalculate the smearing ratio, $S_n$, and determine
the impact on the extracted $R_{np}$ data points. Again, this effect is
largest at large $x$, and can be parameterized as $\delta R_{np} = 0.05
\exp{(-12(1-x))}$.

Combining these systematic uncertainties in quadrature, we find a overall
uncertainty of 0.014, and a correlated uncertainty that grows with $x$ of the
form $\delta R_{np} = 0.11 \exp{(-12(1-x))}$. Figure~\ref{fig:syserr} shows
the extracted values of $R_{np}$ at $Q^2=12$~GeV$^2$, along with the error
band associated with the systematic uncertainties.  Our extracted values,
rebinned by a factor of two and including both statistical and systematic
uncertainties are given in Table~\ref{tab:ratio}.

\subsection{$Q^2$ dependence}

\begin{figure}[htb]
\begin{center}\rotatebox{0.0}{\resizebox{4.2in}{!}{
\includegraphics{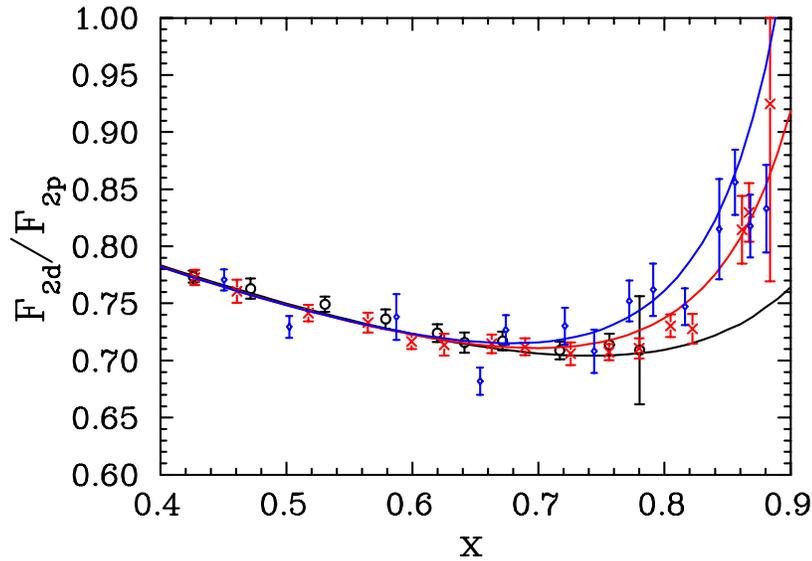}}}
\end{center}
\caption{Ratio of deuteron to proton structure function for
$Q^2$ values from 4--8~GeV$^2$ (black circles), 8--16~GeV$^2$ (red crosses),
and $16-32$~GeV$^2$ (blue diamonds), compared with calculations at $Q^2$=6,12,
and 20~GeV$^2$, using the 7-parameter fit to $R_{np}(\xi)$ shown in
figure~\ref{fig:nprat}.}
\label{fig:qsqdep_x}
\end{figure}

While the analysis is performed at fixed $Q^2$, we fit $R_{np}$ as a function
of $\xi$, with the expectation that the $Q^2$ dependence will be minimal when
taken at fixed $\xi$, as is observed for the proton and deuteron structure
functions.  By assuming that this parameterization is a reasonable
representation of the $Q^2$ dependence of the ratio, we can use this and 
the $Q^2$-dependent fit to the proton structure function~\cite{christy_priv}
to examine the $Q^2$ dependence of the calculated $R_{dp}$.

Figure~\ref{fig:qsqdep_x} shows $R_{dp}$ calculated at $Q^2$=6 (bottom curve),
12, and 20~GeV$^2$ (top curve).  There is a significant $Q^2$ dependence at
large $x$, coming mainly from the $Q^2$ dependence of $F_{2p}$.  We compare
this with the data, now taken in smaller $Q^2$ ranges (4--8, 8--16, and
16--32~GeV$^2$), interpolated to the same $Q^2$ values as the curves.  While
the data at large $x$ is limited in the $Q^2$ range, there is a clear $Q^2$
dependence to the measured ratios which is at least in qualitative agreement
with the result of the calculation based on the extracted $R_{np}(\xi)$.


 \section{Discussion of the Results}

In the deep-inelastic limit, $x^2 \ll Q^2/4m^2$, structure functions are 
related to parton distributions $d(x)$ and $u(x)$. In that limit,
\be
\frac{d(x,Q^2)}{u(x,Q^2)} \approx \frac{4R_{np}(x,Q^2)-1}{4-R_{np}(x,Q^2)}\; ,
\label{eq:doveru}
\ee
plus small contributions from heavier quarks.  For $Q^2=$12~GeV$^2$ this is
expected to be valid for $x \ltorder 0.5$.  Our extraction of $R_{np}(x)$ is
consistent with that calculated from the CTEQ6L~\cite{pumplin02} parton
distributions at all $x$ values, although both extractions have large
uncertainties at the larger $x$ values, where the simple connection to the
quark distributions is expected to break down.

\begin{figure}[htb]
\begin{center}
\includegraphics*[width=3.7in]{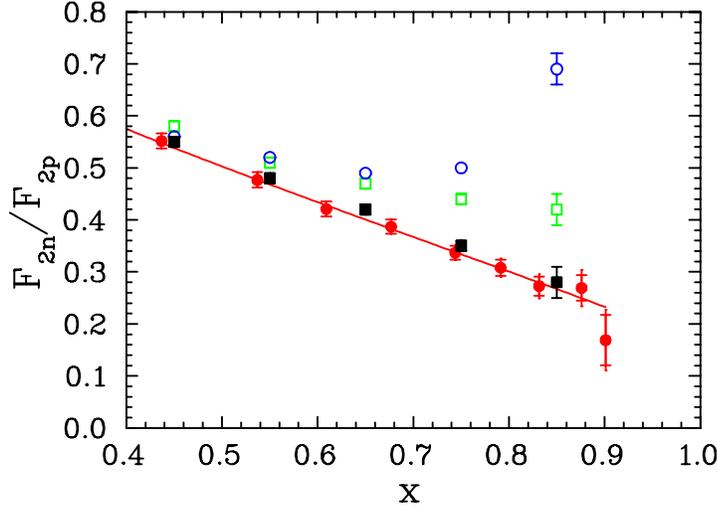}
\end{center}
\caption{Extracted $F_{2n}/F_{2p}$ for $Q^2$=12~GeV$^2$ (solid
circles) compared to previous extractions. The hollow squares are points from
~\cite{melnitchouk96}, while the other points are from~\cite{whitlow92},
using a DIS convolution (solid squares) and using a density-dependent
extrapolation of the EMC effect (hollow circles).}
\label{fig:nprat_compare}
\end{figure}

At large $x$, the quantitative features of the extracted neutron structure
function are limited by the availability of high precision deuteron and proton
data.  Existing measurements provide a reasonable basis for the extraction of
$R_{np}$ with high precision up to $x=0.85$, but no significant constraint for
$x>0.9$. It follows from the proton and deuteron data, and the deuteron
structure assumptions we specified, that the neutron to proton ratio is small
for large $x$, although there is little to constraint the extrapolation to
$x=1$.

In figure~\ref{fig:nprat_compare} we compare the $R_{np}$ ratio obtained in
section~\ref{sec:extraction} to previously published values of $F_{2n}/F_{2p}$.
The solid squares are the results of Whitlow et al.~\cite{whitlow92}, who use
the DIS convolution and the deuteron wavefunction of the Paris
potential~\cite{lacombe81}. The hollow squares are the results
of~\cite{melnitchouk96}, which emphasizes the off-shell effects of the
Buck-Gross~\cite{buck79} spectator representation of the deuteron.

\begin{figure}[htb]
\begin{center}
\includegraphics*[width=3.7in]{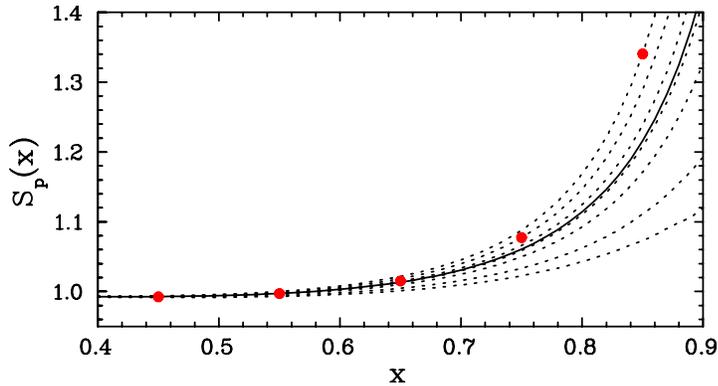}
\end{center}
\caption{The smearing ratio, $S_p$, calculated at several $Q^2$
values. The solid line is 12~GeV$^2$, while the dotted lines correspond to the
$Q^2$ values of the $R_{dp}$ points extracted by Whitlow~\cite{whitlow92}, and
used in previous analysis. The lowest curve corresponds to $Q^2=4.7$~GeV$^2$,
while the upper curve is 23.6~GeV$^2$. The points are placed on the curve
associated with the average $Q^2$ for that $x$ value, which is significantly
different from the 12~GeV$^2$ value for the large $x$ points.}
\label{fig:smearing_qsq}
\end{figure}

There are a few difficulties in comparing these previous results to our
extraction. As mentioned in section~\ref{sec:systematics}, the previous
extractions compared in figure~\ref{fig:nprat_compare} neglect the fact that
the uncertainty in the fits to $F_{2p}$ and the extracted $F_{2n}$ affect the
smearing ratios $S_p$ and $S_n$ (\ref{eq:smearing}), and thus their total
uncertainties are close to our statistical uncertainties, as most of the
additional systematics are neglected.  In addition, the $x$ binning is very
coarse. For $x>0.7$, there is a significant $x$-dependence to $R_{dp}$.  If
the average $x$ of the data in one of the large $x$ bins is shifted from the
center of the bin, placing the point at the central $x$ value will affect the
results.

More important is the strong dependence of the smearing ratios on $Q^2$,
combined with the strong variation of the average $Q^2$ value used for each $x$
bin which affects the results compared in figure~\ref {fig:nprat_compare}.
These extractions used $R_{dp}$ values from~\cite{whitlow92}, where each $x$
value has a different average $Q^2$. The low $x$ points have $\langle Q^2
\rangle$ values below 8~GeV$^2$, while the highest $x$ point has $\langle Q^2
\rangle$=23.6~GeV$^2$. The extraction of~\cite{melnitchouk96} calculates $S_p$
and $S_n$ at $Q^2$=12~GeV$^2$, and uses this for all $x$ values.  However, the
smearing functions have a strong dependence on $Q^2$ at large $x$ values, as
shown for $S_p$ in Figure~\ref{fig:smearing_qsq}. For $x=0.85$, the difference
between $S_p$ at 23.6~GeV$^2$ at 12~GeV$^2$ (the solid line) is 12\%, and so
the proton contribution to $R_{dp}$ should be approximately 0.12 higher, making
$R_{np}$ roughly 0.12 lower.  A precise comparison of the results obtained
using these different models for the deuteron structure contributions will
require a self-consistent treatment of the data in the extraction of $R_{np}$.

If one assumes~\cite{frankfurt81} that the ratio $2F_{2d}/(F_{2p}+F_{2n})$ can
be obtained by scaling the EMC ratio, $F_{2A}/F_{2d}$, then $R_{np}$ can be
extracted from $F_{2d}/F_{2p}$ data without calculating smearing ratios $S_p$
and $S_n$. Such an extraction was performed in~\cite{whitlow92}, yielding the
largest values (hollow circles) for the ratio in
figure~\ref{fig:nprat_compare}. This procedure involves the questionable
assumption that the details of the nucleon momentum distribution in the
deuteron are not important~\cite{melnitchouk00, yang00}. Further, it entirely
neglects the important difference between $S_p$ and $S_n$ by extracting a
universal smearing ratio from the extrapolation of the EMC ratio.

No clear conclusions can be drawn from comparisons of the results of different
extractions unless they take consistent proton and deuteron data as input and
properly treat the $Q^2$ dependence of the data. Given proton and deuteron
data at fixed $Q^2$, the effect of different deuteron structure assumptions
should be investigated. It will be important not to confuse different
structure assumptions with different representations of the same deuteron
structure. Here we assumed that the deuteron state is represented by a vector
in the two-nucleon Hilbert space (tensor product of two single-nucleon Hilbert
spaces), together with the impulse current tensor. The deuteron wave function
which implements these assumptions is an eigenfunction of a mass operator
which implements one-pion exchange and two-pion exchange with some
parameterization of the short range features which affects the high
momentum features of the momentum density $\rho(\vk)$. This is implemented
differently by different potentials which produce high precision fits to
nucleon-nucleon scattering data.

Relations to larger nuclei are important. The impulse assumption with
null-plane kinematics applied here to the deuteron is easily applied to to
larger nuclei, where eigenfunctions of realistic mass operators exist. Ratios
of $^4$He/$^2$H will test the the smearing effects on the average nucleon
structure functions while $^3$He/$^2$H ratios will provide an additional test
of both the nuclear effects and the extracted the neutron structure function
New measurements of these ratios at large $x$ will be available soon from a
recent Jefferson Lab measurement~\cite{arrington07c}.

\section{Conclusion}\label{sec:conclusion}

The relativistic quantum theory of nuclear structure provides the basis for
reliable extraction of neutron structure functions from proton and deuteron
structure functions.  We extract the neutron structure function using an
impulse approximation and an effective two-nucleon mass operator.  No deep
inelastic approximations or quark structure assumptions are involved.
We find that sufficiently accurate interpolations of the proton and deuteron
data to fixed $Q^2$ are essential, and quantitative comparisons to other
extraction procedures are not possible until these extractions include a
proper treatment of the $Q^2$ dependence of the data.

We obtain a
precise extraction of the neutron structure function up to $x \approx 0.85$,
within the context of our deuteron structure approximation.  For $x>0.9$,
there are no data to constrain the calculation, and any extrapolation of the
results to larger $x$ is unreliable.  More data, in particular at large
$x$ values, will be essential to extending the extraction.

\ack

This work was supported by the U. S. Department of Energy, Office of Nuclear
Physics, under contract DE-AC02-06CH11357. The authors would like to thank
M.~E.~Christy for his fits to the proton structure function and for useful
discussions of the data at large $x$, and to W.~Melnitchouk for useful
discussions.  We thank R.~Machleidt for supplying the CD-Bonn  wave functions.

\section*{References}
\bibliographystyle{unsrt_with_abbreviations}
\bibliography{noverp}

\end{document}